\documentclass[a4paper,12pt]{article}
\pdfoutput=1
\usepackage{feynmp-auto,expdlist}
\usepackage{amsmath, amsfonts, amssymb,mathbbol}
\usepackage{graphicx}
\usepackage{enumerate,cancel}
\usepackage{hyperref}
\usepackage{latexsym}
\usepackage{dsfont}
\usepackage{hepnicenames}
\usepackage{enumerate}
\usepackage{soul}
\usepackage[normalem]{ulem}
\usepackage{comment}
\usepackage{feynmp-auto}

\usepackage{units}

\usepackage{soul}

\usepackage{mathrsfs,graphicx,rotating,amsmath,amsfonts,mathtools,booktabs,amssymb,wasysym}
\usepackage{hyperref}\usepackage{slashed}
\usepackage[nosort]{cite}
\usepackage[table,xcdraw,dvipsnames]{xcolor}
\usepackage{bm}
\usepackage{graphicx}
\usepackage{multirow,multicol}
\hypersetup{colorlinks,bookmarksopen,bookmarksnumbered,
	linkcolor=blus,pdfstartview=FitH,urlcolor=rossos,citecolor=verde}
\allowdisplaybreaks

\renewcommand{\[}{\left[}

\newcommand{\mio}[1]{}

\newcommand{\bpm}{\begin{pmatrix}}
\newcommand{\epm}{\end{pmatrix}}

\usepackage{mathrsfs}

\allowdisplaybreaks
\usepackage{multicol}
\usepackage{color}
\definecolor{rosso}{cmyk}{0,1,1,0.4}
\definecolor{rossos}{cmyk}{0,1,1,0.55}
\definecolor{rossoc}{cmyk}{0,1,1,0.2}
\definecolor{blu}{cmyk}{1,1,0,0.3}
\definecolor{blus}{cmyk}{1,1,0,0.6}
\definecolor{bluc}{cmyk}{1,1,0,0.1}
\definecolor{verde}{cmyk}{0.92,0,0.59,0.25}
\definecolor{verdec}{cmyk}{0.92,0,0.59,0.15}
\definecolor{verdes}{cmyk}{0.92,0,0.59,0.4}

\newcommand{\eq}[1]{~{\rm (\ref{eq:#1})}}

\newcommand{\GeV}{\,{\rm GeV}}
\newcommand{\TeV}{\,{\rm TeV}}

\newcommand{\beq}{\begin{equation}}
\newcommand{\eeq}{\end{equation}}

\newcommand{\bea}{\begin{eqnarray}}
\newcommand{\eea}{\end{eqnarray}}
\newcommand{\be}{\begin{equation}}
\newcommand{\ee}{\end{equation}}
\font\tenrsfs=rsfs10 at 12pt
\font\sevenrsfs=rsfs7
\font\fiversfs=rsfs5
\newfam\rsfsfam
\textfont\rsfsfam=\tenrsfs
\scriptfont\rsfsfam=\sevenrsfs
\scriptscriptfont\rsfsfam=\fiversfs
\def\be#1\ee{\begin{equation}#1\end{equation}}
\def\bl#1\el{\begin{align}#1\end{align}}
\def\ba#1\ea{\begin{align*}#1\end{align*}}

\renewenvironment{thebibliography}[1]
{\begin{multicols}{2}[\section*{\refname}]%
		\@mkboth{\MakeUppercase\refname}{\MakeUppercase\refname}%
		\list{\@biblabel{\@arabic\c@enumiv}}%
		{\settowidth\labelwidth{\@biblabel{#1}}%
			\leftmargin\labelwidth
			\advance\leftmargin\labelsep
			\@openbib@code
			\usecounter{enumiv}%
			\let\p@enumiv\@empty
			\renewcommand\theenumiv{\@arabic\c@enumiv}}%
		\sloppy
		\clubpenalty4000
		\@clubpenalty \clubpenalty
		\widowpenalty4000%
		\sfcode`\.\@m}
	{\def\@noitemerr
		{\@latex@warning{Empty `thebibliography' environment}}%
		\endlist\end{multicols}}

\newcommand{\SU}{\,{\rm SU}}

\makeatletter
\font\ital=cmu10

\def\hhref#1{\href{http://arxiv.org/abs/#1}{arXiv:#1}}
\usepackage{xstring}
\newcommand{\hhrefq}[1]{\IfSubStr{#1}{:}{\href{http://inspirehep.net/search?ln=en&ln=en&p=#1&of=hb&action_search=Search&sf=&so=d&rm=&rg=25&sc=0}{InSpire:#1}}{\hhref{#1}}}

\def\art{\@ifnextchar[{\eart}{\oart}}
\def\eart[#1]#2#3#4#5#6{{\rm #2}, {\em #3 \bf #4} {\rm (#6) #5} ({\em #1})}
\def\article{\@ifnextchar[{\earticle}{\oarticle}}
\def\oarticle#1#2#3#4#5#6{{\rm #1}, {\ital `#6'}, {\rm #2 #3 (#5) #4}}
\def\earticle[#1]#2#3#4#5#6#7{{\rm #2}, {\ital `#7'}, {\rm #3 #4 (#6) #5}  [\hhrefq{#1}]}
\def\hepart[#1]#2{{\rm #2, \sl#1}}
\def\heparticle[#1]#2#3{#2, {\ital `#3'} [\hhrefq{#1}]}
\newcommand{\doi}[1]{\href{http://dx.doi.org/#1}{[link]}}

\newcommand{\hhrefqq}[1]{\IfBeginWith{#1}{10.}{\href{https://doi.org/#1}{doi:#1}}{\hhrefq{#1}}}

\renewenvironment{thebibliography}[1]
{\begin{multicols}{2}[\section*{\refname}]%
		\@mkboth{\MakeUppercase\refname}{\MakeUppercase\refname}%
		\list{\@biblabel{\@arabic\c@enumiv}}%
		{\settowidth\labelwidth{\@biblabel{#1}}%
			\leftmargin\labelwidth
			\advance\leftmargin\labelsep
			\@openbib@code
			\usecounter{enumiv}%
			\let\p@enumiv\@empty
			\renewcommand\theenumiv{\@arabic\c@enumiv}}%
		\sloppy
		\clubpenalty4000
		\@clubpenalty \clubpenalty
		\widowpenalty4000%
		\sfcode`\.\@m}
	{\renewcommand{\@noitemerr}
		{\@latex@warning{Empty `thebibliography' environment}}%
		\endlist\end{multicols}}

\newcounter{alphaequation}[equation]
\renewcommand{\thealphaequation}{\theequation\hbox to
	0.6em{\hfil\alph{alphaequation}\hfil}}
\definecolor{Gray}{gray}{0.95}

\usepackage{tocloft}
\begin{document}
\thispagestyle{empty}
\begin{center}  
{\LARGE\bf\color{rossos} Is nature natural?} \\[3ex]
 {\bf Alessandro Strumia}  \\[1ex]
{\it Dipartimento di Fisica, Universit\`a di Pisa, Italia}\\[3ex]
{\large\bf Abstract}
\begin{quote}\large
I summarize the apparent relevance of anthropic selection in understanding 
unnatural features of the Standard Model, and the consequent implications for theory.
This topic, pioneered by Weinberg, could make him the last giant of physics.

\end{quote}

\noindent

\bigskip
{\small\color{blus} Contribution to ``A Master's Journey through the Two Standard Models: \\
A volume dedicated to the Memory of Steven Weinberg''.}\\[2ex]

\end{center}
\setcounter{page}{1}

\bigskip\bigskip


\noindent
Weinberg proposed so many important ideas that 
one can go beyond remembering his well-celebrated achievements,
and recognise his greatness by discussing unresolved issues.
Unlike successes, these issues remain open today.

\smallskip

Weinberg started working when fundamental theory was about understanding unexplained experimental findings.
When Weinberg proposed ``a model of leptons''~\cite{Weinberg:1967tq} 
nobody cared much for a few years.
Now the situation changed so much that we have
the opposite problem: it's difficult to understand the difficulties he overcome.
Some key progress in physics comes from simple ideas that everybody ignored 
because nobody dared to think differently.
The SM is based on Quantum Field Theory.
When Weinberg wrote his only model, alternative theories such as ``nuclear democracy''
were considered more promising.
Today QFT is recognised to be the fundamental language that describes physics,
at least up to the quantum gravity scale.
Choosing and developing the right language was a key Weinberg achievement.

\smallskip

Unexpectedly, the Standard Model worked so well 
that nothing new was discovered in the next 50 years, up to now.
Without having experimental discoveries to be understood,
fundamental physics drifted to searching/guessing new physics beyond the SM,
and to trying to understand puzzling theoretical aspects of the SM.
Weinberg and everybody else got stuck. 
He said: {\em ``I don't know how much elementary particle physics can improve over what we have now.  [...] 
I hope it doesn't just stop where it is now. Because I don't find this entirely satisfying''}~\cite{LiM}.

These words by Weinberg come from an interview about naturalness, the topic of this article.

\medskip

Many theorists argued that the Higgs boson should be accompanied by weak-scale
new physics able of keeping its mass $M_h$ naturally much smaller than the Planck scale.
Weak-scale supersymmetry seemed the most plausible solution to the hierarchy problem.
Weinberg wrote, in a book about supersymmetry,  that
{\em ``because of the intrinsic attractiveness of supersymmetry and the possibility it offers of
resolving the hierarchy problem, I and many other physicists are reasonably confident that supersymmetry will be found''}, mentioning the LHC.

\smallskip

According to Weinberg, {\em ``that's been a huge disappointment''}~\cite{LiM}.
Although various theorists expressed similar thoughts, Weinberg's perspective carries particular significance.
By finding  the Higgs boson with mass $M_h \approx 125.1\GeV$ not accompanied by any new physics,
the Large Hadron Collider had confirmed once again the Standardissimo Model, proposed by Weinberg.
This situation is theoretically interesting because it contradicts conventional ideas about naturalness.

\smallskip

Weinberg dared to think outside such conventional ideas.
Assuming instead dynamical electro-weak symmetry breaking within the SM,
Gildener and Weinberg predicted a Higgs mass $M_h \approx 7\GeV$~\cite{Gildener:1976ih}.
The general idea of dynamical symmetry breaking was
pioneered by Sidney Coleman and Erick Weinberg~\cite{Coleman:1973jx}.
It remains applicable to the electro-weak symmetry:
more complicated models of new physics
avoid the unsuccessful prediction for the Higgs mass.
Rather than discussing such attempts, 
I go for what seems to me the most plausible direction: anthropic selection in a multiverse.
In this way, I follow Weinberg golden lessons~\cite{Wein4gold}
\begin{itemize}
\item  go for the messes, aim for rough water;
\item forgive yourself for wasting time;
\item learn something about the history of science.
\end{itemize}

%

\subsection*{The naturalness puzzles}
Oversimplifying a bit, we observe three main vastly different scales in nature: 
the Planck scale $M_{\rm Pl}$ of quantum gravity;
the weak scale $v$ set by the Higgs mass $M_h^2 = V''(v)$;
the vacuum energy scale $V(v)$. 
The latter two scales are related to the Higgs potential $V(h)$ and data tell
\beq \label{eq:VV''}
V(v) \sim 10^{-123} M_{\rm Pl}^4,\qquad
V''(v) \sim 10^{-34}M_{\rm Pl}^2.\eeq
Attempts of formulating and testing
theories that naturally explain  the huge hierarchy of scales lead to the following
current situation:
\begin{itemize}
\item[1)] Nobody ever found a theory of new physics that keeps 
the SM vacuum energy  naturally small.
Cosmological data are compatible with a cosmological constant $V(v)$.

\item[2)] Some theories of new physics at the weak scale
that could have kept the Higgs mass naturally small have been found, such as
supersymmetry.
However, LHC searches found just the SM Higgs boson.
\end{itemize}
A possible lesson is that both $V(v)$ and $V''(v)$ are unnaturally small.

A century ago many theorists believed that the
 \ae{}ther theory was obviously true, reasoning that `there is a wave, so there must be a medium'.
When  the  
Michelson-Morley null result cast doubts on the  \ae{}ther, many 
theorists tried saving the  \ae{}ther (e.g.\ by suggesting it might be dragged by the Earth),
before accepting an interesting alternative idea, relativity.

\smallskip

A similar story is happening with Higgs naturalness.
When LEP cast doubts on this theory~\cite{hep-ph/0007265}
many theorists proposed special models.
After the LHC, research in this direction is waning and
alternative understandings of the naturalness puzzles are being explored.
Possible ideas include cosmological adjustment mechanisms, or
the absence of new heavy physics significantly coupled to the Higgs.
However the most plausible understanding of the disappointing null results 
remains the least discussed, because it appears more disappointing than the null result:
anthropic selection in a multiverse.

\subsection*{Anthropic selection}
The idea is that
the seemingly unnatural values in eq.\eq{VV''} are indeed unnatural:
they arise via accidental cancellations among large contributions,
that happen with a small probability estimated as $\wp \sim 10^{-34}\cdot 10^{-123}$.
Nevertheless rare events with probability $\wp \ll 1$ happen 
if some mechanism allows at least $1/\wp \sim 10^{155}$ trials. 
A plausible mechanism will be discussed in the next section.
Independently of the mechanism, the general idea behind anthropic selection is that
we live in a rare atypical `low-energy'  vacuum where 
complex `life'  is made possible by:
\begin{itemize}
\item[1)] a tiny $V(v)$, meaning a quiet Minkowski space-time 
(while a big  $V>0$ would have lead to fast de Sitter expansion,
and a big $V<0$ to a fast anti-de Sitter crunch);
\item[2)]  long-lived particles with low mass $m \sim yv $, where $v$ is the weak scale
and $y$ are dimensionless couplings.
Weinberg view of the SM as an Effective Field Theory explains the near-stability
of the proton as due to the smallness of the weak scale
compared to the scale that suppresses non-renormalizable effective operators that violate baryon number.
Gravity is one non-renormalizable effect that was observed because gravitational effects accumulate in bodies with large mass.
A small $m \ll M_{\rm Pl}$  generically allows macroscopic bodies made of 
up to $N \lesssim (M_{\rm Pl}/m)^3 \gg 1$ particles.
\end{itemize}
If the above generic idea is true,
unnaturally small scales should be just mildly smaller than what is anthropically needed.
It should be possible to go beyond the qualitative arguments in 1) and 2) and
find  precise anthropic boundaries that prevent the weak scale and the vacuum energy
to be significantly larger than their observed values.
Such boundaries have been indeed claimed.
\begin{itemize}
\item[1)] Concerning the vacuum energy, the first key step was taken by Weinberg~\cite{Weinberg:1987dv}.
By combining cosmology with anthropic selection, Weinberg showed
that a too large vacuum energy would have prevented galaxy formation.
He proposed that $V/\delta^3$
(where $\delta \approx 10^{-5}$ is the amplitude of primordial inhomogeneities)
must be small enough due to anthropic selection,
but does not  need to vanish.
A cosmological constant somehow below the Weinberg anthropic bound has later been observed.
\end{itemize}
Weinberg never applied anthropic selection to understand the Higgs naturalness problem.
His 2000 book about supersymmetry mentioned the naturalness motivation,
without mentioning that an alternative anthropic understanding had been proposed in~\cite{hep-ph/9707380}.
These authors observed that the SM predicts $Z \approx 100$ different nuclei
not because of some nice argument (no simple understanding such as
$Z\sim 1/\alpha$ holds),
but because of a numerical coincidence that makes nuclear physics complicated.

\begin{itemize}
\item[2)] 
The proton is the lightest quasi-stable baryon, despite not being the more neutral baryon nor being the baryon made of the lighter quarks. 
QCD long-range forces attract protons to neutrons, so both are needed to form nuclei.
Nuclei exist because neutrons, being just slightly heavier than protons, become stable within nuclei. 
This peculiar situation is realized within the SM because two different effects happen to be comparable,
despite originating from different scales:
\beq m_n-m_p=
\underbrace{ {\cal O}(\alpha_{\rm em} \Lambda_{\rm QCD})}_{-0.10\%}
~~ +
\underbrace{ (y_d-y_u) v}
_{0.25\%}\eeq
where the first term is the electromagnetic energy that arises because the proton is charged,
and the second is the $uud$ vs $udd$ contribution due to quark masses.
Changing the Higgs vev $v$ by a factor of few removes this coincidence,
making nuclear physics simple (just one or two nuclei),
and thereby chemistry and `life' impossible.
\end{itemize}
However, the argument  in~\cite{hep-ph/9707380} only sets an anthropic bound on $yv$. 
So it is not enough to explain the unnatural smallness of $v$,
as a more natural SM-like theory could have achieved the same $m_n - m_p$
with a larger $v$ times a smaller $y$ (small Yukawa couplings are technically natural).
For example, a weak-less theory with $v \sim M_{\rm Pl}$ would be
natural.
In order to understand anthropically the smallness of $v$,
one needs a direct anthropic bound on $v$, that controls weak interactions.
Weak interactions enter two processes of big cosmological relevance.
In both cases the non-trivial SM outcome is crucially related to the following numerical
consequence:
\beq \label{eq:van}
v \sim \Lambda_{\rm QCD}^{3/4} M_{\rm Pl}^{1/4} \sim \TeV .\eeq
\begin{itemize}
\item[$2')$] The first is Big Bang Nucleosynthesis (BBN).
The numerical coincidence of eq.\eq{van} implies that
the neutrino decoupling temperature $T_{\nu\rm dec}$
is comparable to the the proton-neutron mass difference,
and that the neutron life-time $\tau_n$
is comparable to the age of the Universe at BBN time, $t_{\rm BBN}$.
This coincidence makes the first 3 minutes of the Universe so interesting that Weinberg described what happened in a book~\cite{3min}.
The final order-one ratio 
between the number of neutrons $N_n$ and of protons $N_p$ is
\beq \frac{N_n}{N_p} \approx \exp\left(-\frac{m_n-m_p}{T_{\nu\rm dec}} -\frac{ t_{\rm BBN}}{\tau_n}\right)\approx \frac17.
\eeq
However, this means that $N_n/N_p$ is exponentially sensitive to such coincidences only if $v$ is lowered,
without leading to a BBN anthropic upper bound on $v$~\cite{1409.0551}.

\item[$2'')$]  Core-collapse supernova explosions are complicated because
the gravitational time-scale of the collapse is comparable to the escape times of neutrinos.
Actually,  two different escape times, from the surface and from the volume,
get comparable to $\tau_{\rm grav}$ when eq.\eq{van} holds:
\beq \tau_{\rm grav} \sim \frac{M_{\rm Pl}}{\Lambda_{\rm QCD}^2} \sim{\rm sec},\qquad
\tau_{\rm volume} \sim \frac{\Lambda_{\rm QCD} M_{\rm Pl}^2}{v^4},\qquad
\tau_{\rm surface} \sim \frac{M_{\rm Pl}^{3/2}}{v^2 \Lambda_{\rm QCD}^{1/2}}.\eeq
Difficult computations suggest that SN explosions happen
thanks to neutrinos that, at the right moment, come out and push the material that surrounds the core.
SN explosions could have anthropic relevance because  
some  intermediate-mass  elements  (O, F, Ne, Na, Mg, Al and possibly N, Cl, K, Ca) 
seem produced almost exclusively in core collapse SN~\cite{Woosley:1995ip,Astroper} and
seem needed for `life'~\cite{Barrow:1988yia}.
\end{itemize}
This understanding of the hierarchy problem has weak aspects, which could be better clarified. 
For example~\cite{1906.00986} could only simulate 
SN in spherical approximation, while order one factors are so critical that
a-sphericity seems critical to achieve SN explosions.

\subsection*{Anthropic theory}\label{anth}
If the anthropic selection argument is correct, what are its consequences for theory?
The mechanism can be realized in different ways.
One way is fully consistent with our current understanding of theory:
Unfortunately, this way appears to be also the least testable:
having many separate vacua.
Big quantum fluctuations during inflation can `populate' different vacua 
in different regions of the `multiverse' that can become cosmologically large and loose causal contact with each other.

A landscape of many vacua might arise in string compactifications~\cite{hep-th/0302219}.
Furthermore, it is compatible with an ordinary Quantum Field Theory with enough ordinary scalars and an ordinary potential.
This allows a large enough number $\gg 10^{150}$ of different local minima.
Each minimum corresponds to a different set of particles with different masses,
because in QFT particles are small fluctuations around local minima.

Some rare vacua accidentally have small energy density $V(v)$.
In some rare vacua the scalar potential contains one small second derivative,
corresponding to a weak-scale Higgs boson.
This double coincidence allows for a doubly-rare vacuum where complex `life' is possible.

\medskip

This scenario can be realised in a renormalizable QFT
where scalar potentials $V$ are polynomials of quartic order.
Counting the number of minima of this kind of potential is not easy.
Naively, $V$ can have up to $3^N$ extrema. 
Naively, each extremum can be a minimum with probability $2^{-N}$,
as the $N$ eigenvalues of the matrix of second derivatives must be all positive.
So maybe a `typical' quartic potential has many $\sim (3/2)^N$ minima at large $N$.
This needs to be so large that counting the minima and computing their properties
(such as the vacuum energy and particle masses, to see if they are accidentally small)
cannot be done by numerical methods.

A simple analytic argument holds in a special case:
$N$ independent scalars, corresponding to a potential
$V = \sum_{i=1}^N V_i (\phi_i)$ with 2 minima in each scalar $h_i$.
So this $V$ has $2^N$ minima, and their vacuum energies follow a Gaussian distribution~\cite{hep-th/0501082}.
In the above toy example, the scalar masses are nearly the same in all vacua.
Anthropic selection needs $\gg 10^{34}$ different values of $M_h^2$.
This can be achieved by considering a more complicated controllable example,
with an approximate $\mathbb{Z}_2^N$ symmetry that flips the sign of each scalar.
The resulting potential
\beq \label{eq:VZ2N}
V =  V_0-\frac12 M_i^2 h_i^2 +\frac14 \lambda_{ij} h_i^2 h_j^2 + \epsilon V_{\rm odd}\eeq
still has $2^N$ vacua, that persist if the symmetry-breaking $V_{\rm odd}$
is suppressed by a small enough $\epsilon$.
This is enough to get minima with different scalar masses and known distribution~\cite{1911.01441}.


\medskip

QFT implies that a Minkowski vacuum can spontaneously 
tunnel into any of the $\sim 2^N$ vacua with lower negative vacuum energy.
So, does the lucky SM vacuum survive up to cosmologically large times?
One might worry that the decay rate is enhanced by the number of final states $\sim 2^N$, 
but actually just $N$ bounces exist (or dominate) in the example of eq.\eq{VZ2N}.
Then, a typical vacuum is stable enough provided that quartic couplings are not too large, $|\lambda_{ij}|\lesssim 1$.

We are however interested in special vacua, featuring an accidentally light scalar $h$.
Such vacua can be highly unstable.
Indeed the potential can be Taylor-expanded for small $h$ as
\beq \label{eq:Vh}
V(h) \simeq   V(v)+ V'(v) h + \frac{V''(v)}{2} h^2 + \frac{V'''(v)}{6} h^3 + \frac{V''''(v)}{24} h^4 +\cdots\eeq
where $V(v)$ is accidentally small, $V'(v)$ vanishes at the minimum,
$V''(v)$ is accidentally small, but $V'''(v)$ is of the order of the large (Planckian?) energy scale.
Eq.\eq{Vh} descrives a tiny potential barrier, leading to a fast vacuum tunneling.

\medskip

This problem is avoided by the SM Higgs doublet $H = (0,h/\sqrt{2})$, as $\SU(2)_L$ group theory forbids cubic $H^3$ terms.
Other representations of other groups would have been problematic.
The Higgs seems to follow what suggested by multiverse considerations.
Furthermore, the special $\SU(2)_L$ group theory property $2 \sim 2^*$ allows 
one Higgs to have multiple Yukawa interactions.
It can thereby give mass to all fermions (charged and neutral leptons, up-type and down-type quarks) 
as in the Standard Model
\beq  y_E ~  ELH^* +y_D DQH^*  +  y_U\,  UQH + (LH)^2/2\Lambda.\eeq
With some optimism, these aspects of the SM group theory can be seen as anthropic predictions.


\subsection*{Conclusions}
Weinberg pioneered the use of anthropic selection as a possible interpretation of unnatural features.
When he studied the vacuum energy,
the mainstream approach was exploring how something like broken supersymmetry could preserve the vanishing of the energy of the unique vacuum.
At that time cosmology was not yet considered  as a serious science.
Nevertheless, Weinberg  combined cosmology with anthropic selection.
His innovative thinking likely sparked some critical comments, as he later remarked: ``{\em a physicist talking about the anthropic principle
runs the same risk as a cleric talking about pornography:
no matter how much you say you are against it,
some people will think you are a little too interested''}.

\smallskip

Weinberg remained ambivalent about anthropic selection.
On the one hand, he recognised that
{\em ``it might be that the correct theory allows a great number of different big bangs''}~\cite{LiM}.
On the one hand, {\em ``the best we can hope for is a theory that is unique in the sense that it is the only mathematically consistent theory that is rich, that has lots and lots of phenomena, and in particular that makes it possible for life to arise''}~\cite{LiM}.
Weinberg motivated this traditional canon of a beautiful theory 
by arguing that it was shaped by past successes:
{\em  ``gradually our sense of beauty changes in a way that is produced by our experience''}.
Then, current experience could change our sense of beauty, until
we learn how to stop worrying about naturalness and strangelove anthropic selection.

Anthropic selection would make further scientific progress difficult.
If the anthropic understanding is true, if we already found all the particles that exist at the weak scale,
if experiments will keep finding
`everything in the SM, nothing outside the SM, nothing against the SM',
then Weinberg might literally be the last giant of physics.


\end{document}